\documentclass[twocolumn,english,aps,prl,10pt,superscriptaddress,showpacs]{revtex4}
\usepackage[T1]{fontenc}
\usepackage{amsmath,graphicx,amssymb,epsfig,babel,dsfont}
\usepackage{color}
\usepackage{float}

\renewcommand{\Im}{{\rm Im}}

\begin{document}

\title{Multi-tip near-field scanning thermal microscopy}

\author{Philippe Ben-Abdallah}
\email{pba@institutoptique.fr} 
\affiliation{Laboratoire Charles Fabry, UMR 8501, Institut d'Optique, CNRS, Universit\'{e} Paris-Saclay,
2 Avenue Augustin Fresnel, 91127 Palaiseau Cedex, France}
%\affiliation{Universit\'{e} de Sherbrooke, Department of Mechanical Engineering, Sherbrooke, PQ J1K 2R1, Canada}

%\author{Riccardo Messina}
%\affiliation{Laboratoire Charles Fabry, UMR 8501, Institut d'Optique, CNRS, Universit\'{e} Paris-Saclay,
%2 Avenue Augustin Fresnel, 91127 Palaiseau Cedex, France}

%\date{\today}


\begin{abstract}
A theory is presented to describe the heat-flux radiated in near-field regime by a set of interacting nanoemitters held at different temperatures in vacuum or above a solid surface. We show that this thermal energy can be focused and even amplified  in spots that are much smaller than those obtained with a single thermal source. We also demonstrate the possibility to locally pump heat using specific geometrical configurations. These many body effects pave the way to a multi-tip near-field scanning thermal microscopy which could find broad applications in the fields of nanoscale thermal management, heat-assisted data recording, nanoscale thermal imaging, heat capacity measurements and infrared spectroscopy of nano-objects.
\end{abstract}

\maketitle

Heat flux focusing radiated by a hot object at temperature $T$ is limited in far-field regime by the diffraction to  $\lambda_{th}/2$ where $\lambda_{th}=\hbar c/k_B T$ is the thermal wavelength associated to this source.  However at subwavelength distance from the source the situation can radically change. The near-field scanning themal microscope (SThM)~\cite{De Wilde,Achim,Raschke,Hillenbrand,Weng} which is the non-contact version of conventional scanning thermal microscope (STM)~\cite{Williams,Majumdar} can achieve a local  heating at submicrometric scale with heat sources close to the ambiant temperature using the tunneling of non-radiative thermal photons (i.e. evanescent waves). This near-field technology is among others the current paradigm of hard-disk-drive writing technology which is based on the so called heat-assisted magnetic recording~\cite{Challener,Stipe}. In this technique a tiny surface area of a magnetic material is heated up in near-field regime to raise the material temperature close to its Curie temperature, thus demagnetizing it locally. To store a high density of magnetic bit the hot spot area should be ideally reduced as close as possible as the superparamagnetic limit beyond which the bits become unstable due to thermal fluctuations. The typical size to observe superparamagnetism in usual magnetic materials is typically below domains of 20 nm side. However the radiative heat focusing by a conventional scanning probe microscope (SPM) is  limited by the emission pattern in near-field regime of its tip.

\begin{figure}
\includegraphics[scale=0.3]{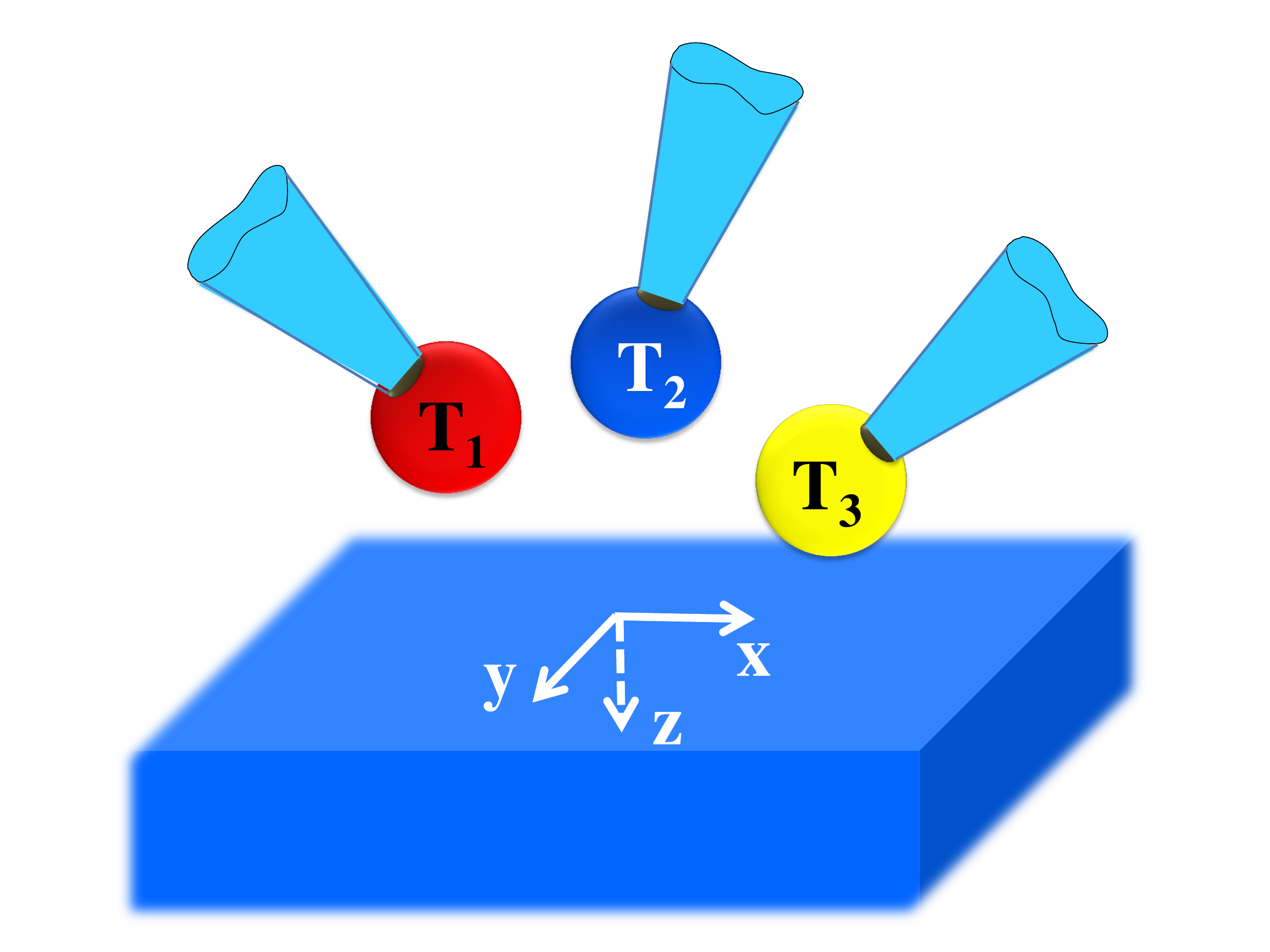}
\caption{Schematic of a multi-tip SThM platform. Nanospheres (thermal emitters) are grafted on single SPM tips. Their temperature and position are individually controlled.
\label{multi_tips_SthM}}
\end{figure}

In the present Letter we theoretically investigate the electromagnetic field radiated by a multi-tip SThM setup (Fig.1) which consists in several SPM units with tips which are individually positioned using nanopositioning systems based, for instance, on differential interferometry between the tips and thermally controled  with thermocouple junctions inserted inside each tip associated to loops of feedback control. Each single source at temperature $T_i$ is a broad band emitter which radiates over a spectral range defined by the Planck's distribution function at temperature $T_i$, its maximum being located at the Wien's wavelength $\lambda_{Wi}(\mu m)=2898/T_i(K)$.  Hence for a set of $N$ emitters at different temperatures $T_1<...<T_N$ we can define $N$ corresponding Wien's wavelengths $\lambda_{WN}<...<\lambda_{W1}$ which set the spectral range $[\alpha \lambda_{WN}, \beta\lambda_{W1}]$ (typically $\alpha\sim 0.1$ and $\beta\sim 100$ ) within which all heat  exchanges occur.  We demonstrate that the heat flux radiates by these thermal emitters can be focused and amplified into spots of much smaller area than with a single tip setup. Moreover we show that this focusing can be achieved up to surface area close to the superparamagnetic limit. Also we demonstrate with sharp geometric configurations  the existence of a near-field heat pumping effect. To highlight those effects and to introduce the basic principles that drive the thermal emission in multi-tip SThM platform we consider elongated SPM tips with glass nanospheres ($R=20 nm$) attached at their apex. With temperatures close to the ambiant temperature the tips radii are extremely small compared with the smallest thermal wavelength so that they act more like dipoles than macroscopic spheres. Hence the multi-tip SThM  platform can be modeled by simple radiating dipoles provide their separation distance  is large enough to make the contribution of multipoles negligeable. Typically this condition is satisfied ~\cite{pbaPRB2008,Rubi,Becerril} when the separation distances (center to center) is  larger than $3R$. 
Under these assumptions the optical behavior of any arbitrary multi-tip setup composed by $N$ tips can be described by the interaction of fluctuating dipoles associated to each nanoemitter. 
The heat flux radiated through an oriented surface by such a set of emitters held at different  temperature $T_i$ can be calculated  from the statistical averaging of Poynting vector spectrum
\begin{equation}
 <\bold{\widetilde{S}}(\bold{r},\omega)>=2 Re<\bold{E}(\bold{r},\omega)\times \bold{H^*}(\bold{r},\omega)>,
\label{Poynting}
\end{equation}

\begin{figure}%[h!]
\includegraphics[scale=0.3]{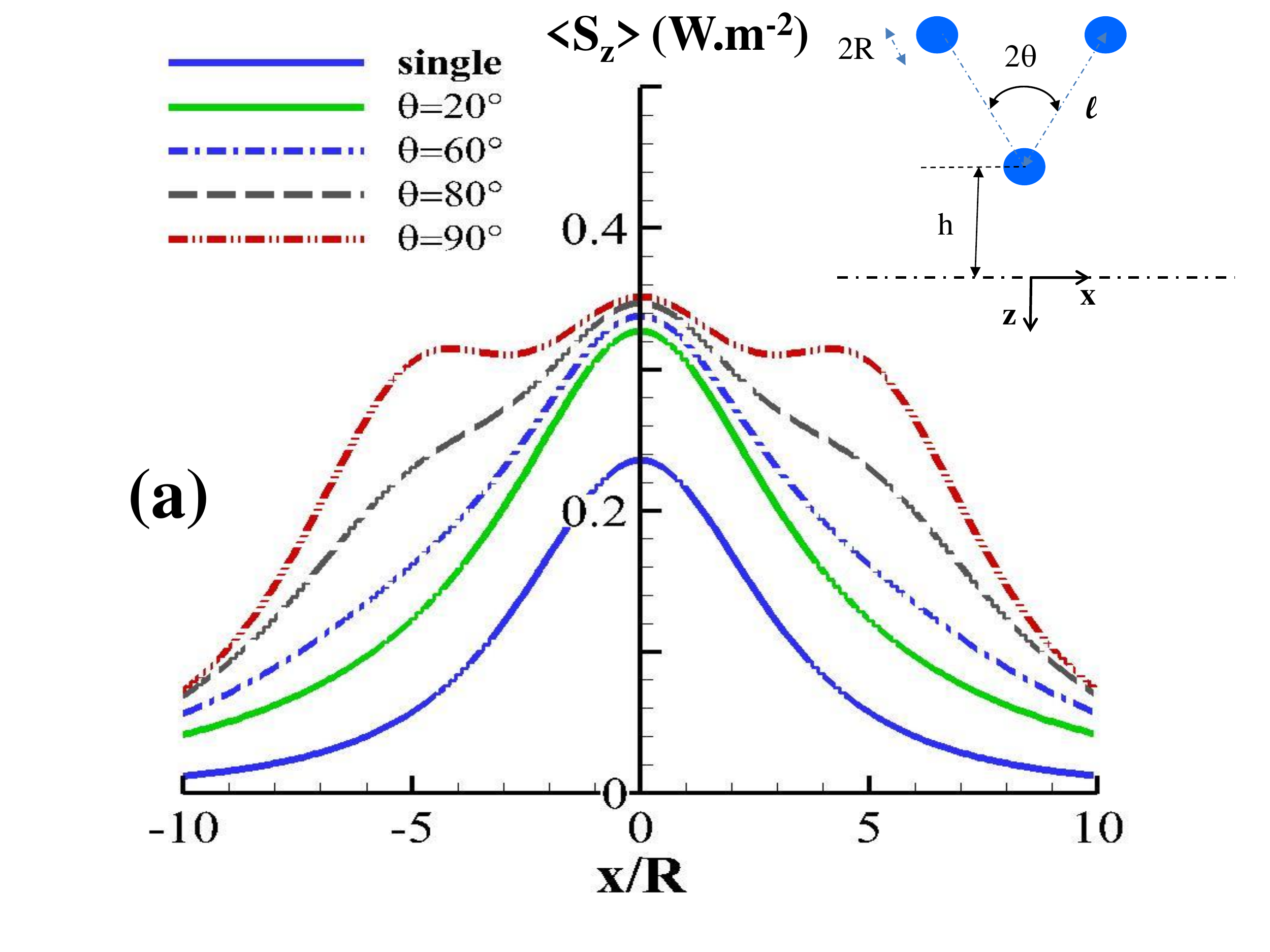}
\includegraphics[scale=0.3]{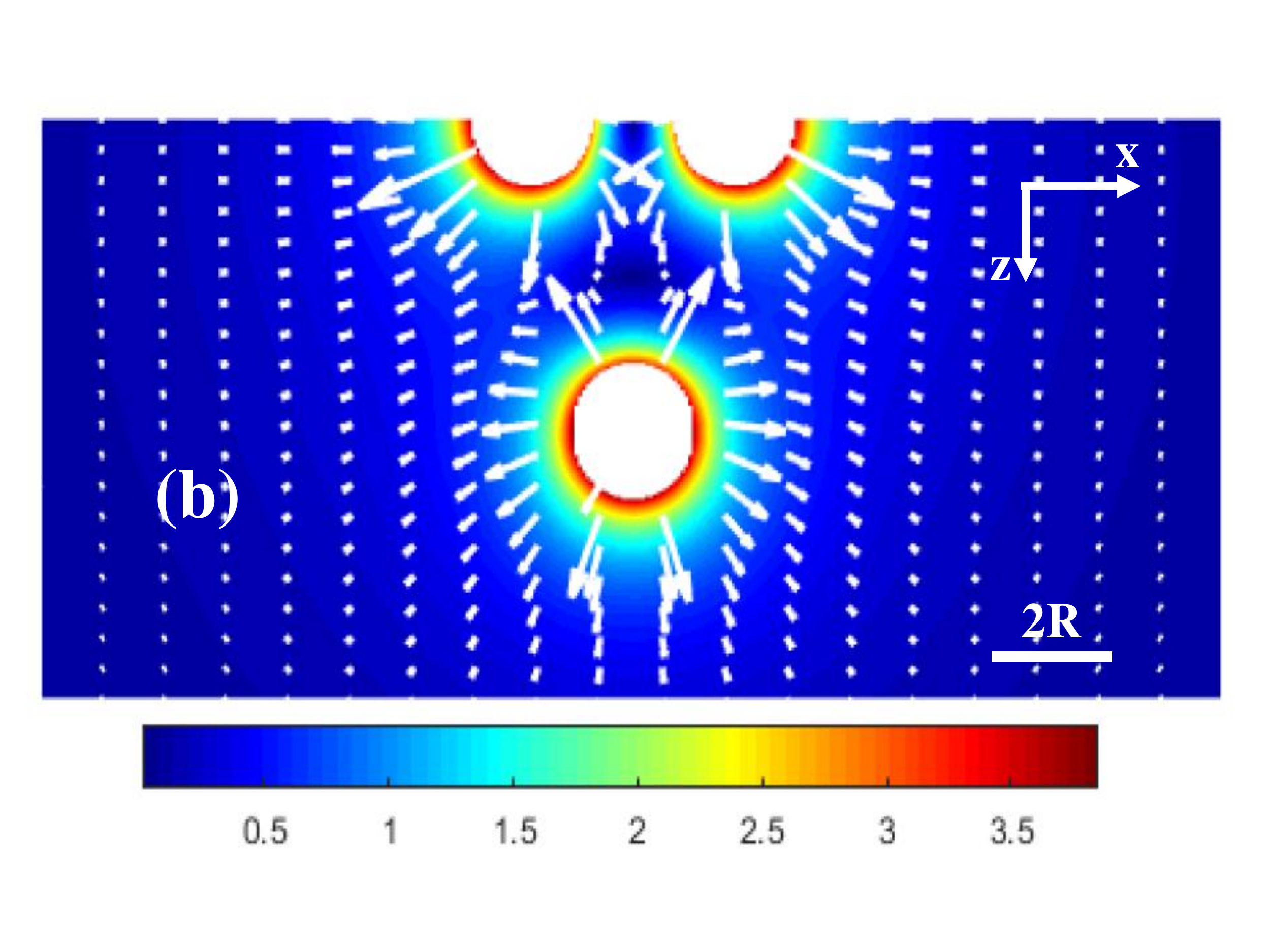}
\caption{(a) Normal component $<S_z>$ of Poynting vector radiated through the surface $z=0$ by a three-tips SThM setup made with nanoemitters in glass~\cite{Palik98} held at the same temperature for different angular opening ($T=300K$, $R=20 nm$, $h=4R$ and $l=5R$). The blue continue line corresponds to the case of a single emitter located at $z=-h$. (b) Magnitude of Poynting vector and vector field in the $(x,z)$ plane radiated by a multi-tip setup of angular opening  $\theta=20^{\circ}$.
\label{focus_same_temp}}
\end{figure}

 where the electric and magnetic field can be related to local fluctuating dipoles $\bold{p_i}^{fluc}$ by the following relations~\cite{PBAEtAl2011}
\begin{equation}
 \bold{E}(\bold{r},\omega)=\omega^2\mu_0 \underset{i=1}{\overset{N}{\sum}}\mathds{G}^{EE}(\bold{r},\bold{r_i})\bold{p_i}^{fluc},
\label{Electric}
\end{equation}

\begin{equation}
 \bold{H}(\bold{r},\omega)=-i\omega \underset{i=1}{\overset{N}{\sum}}\mathds{G}^{HE}(\bold{r},\bold{r_i})\bold{p_i}^{fluc},
\label{Magnetic}
\end{equation}
where $\mathds{G}^{EE}$ and $\mathds{G}^{HE}$ are the full electric and magnetic dyadic Green tensors at the frequency $\omega$ which take into account all many body interactions and eventually interactions with an interface~\cite{SupplMat}. Hence, using the fluctuation dissispation theorem~\cite{Callen}
\begin{equation}
\langle p^{fluc}_{i,\alpha}p_{j,\beta}^{fluc*}\rangle=2\frac{\epsilon_{0}}{\omega}\Im(\alpha_{i})\Theta(\omega,T_{i})\delta_{ij}\delta_{\alpha\beta},
\label{FDT}
\end{equation}
 with $\alpha_{i}$  the polarizability associated to $i^{th}$ dipole), the spectrum of Poynting components reads

\begin{figure}%[h!]
\includegraphics[scale=0.3]{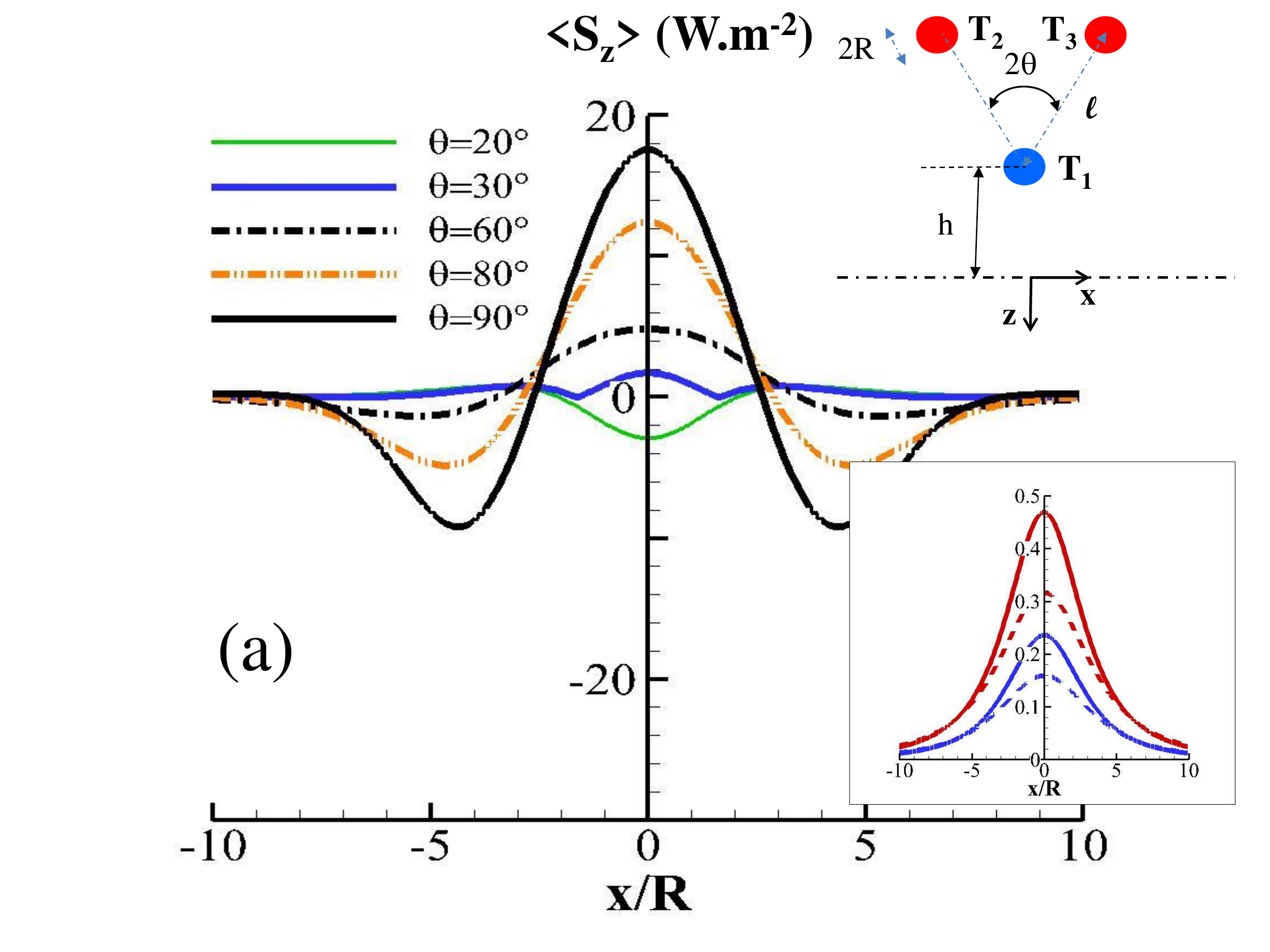}
\includegraphics[scale=0.3]{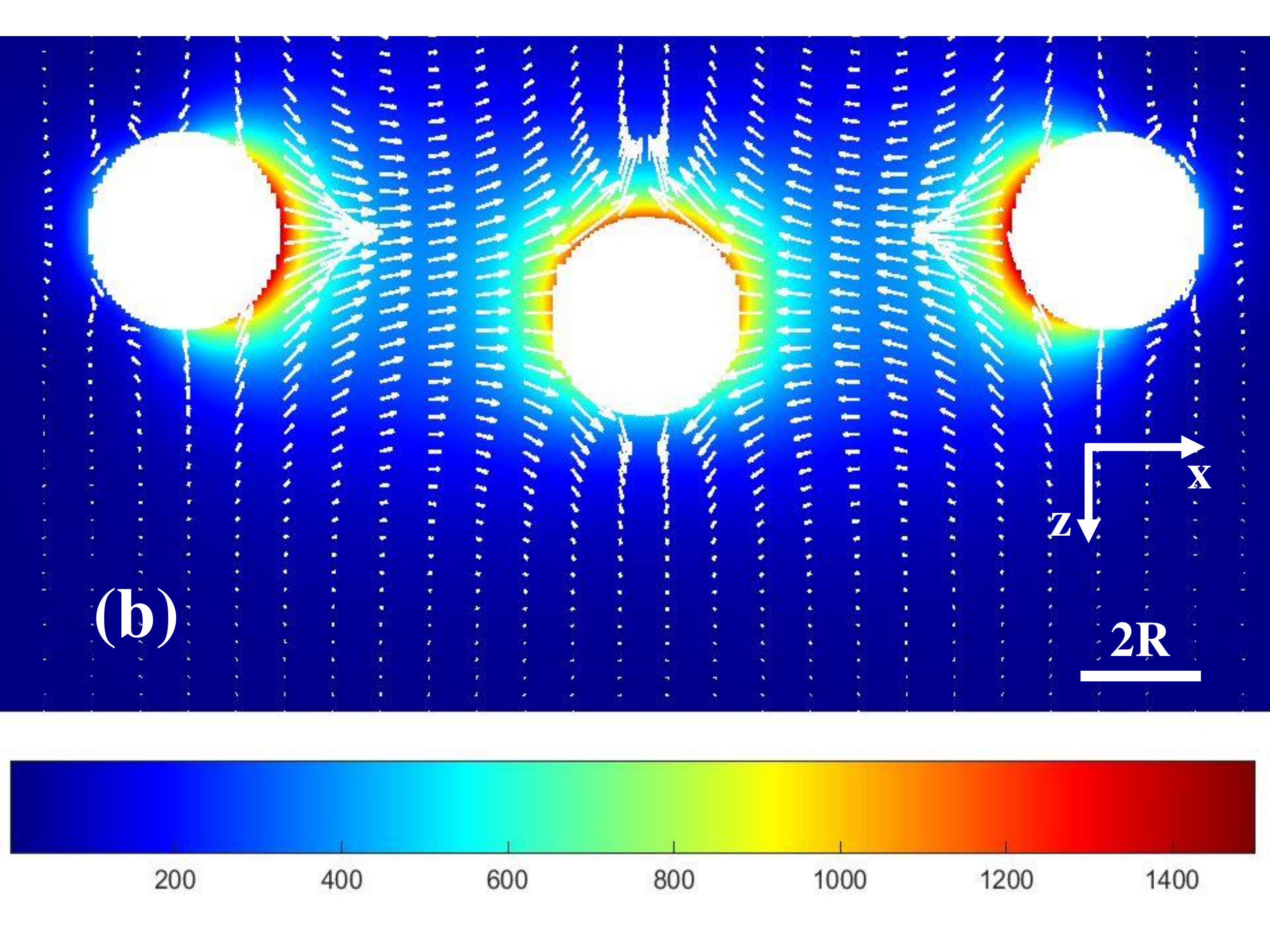}
\includegraphics[scale=0.3]{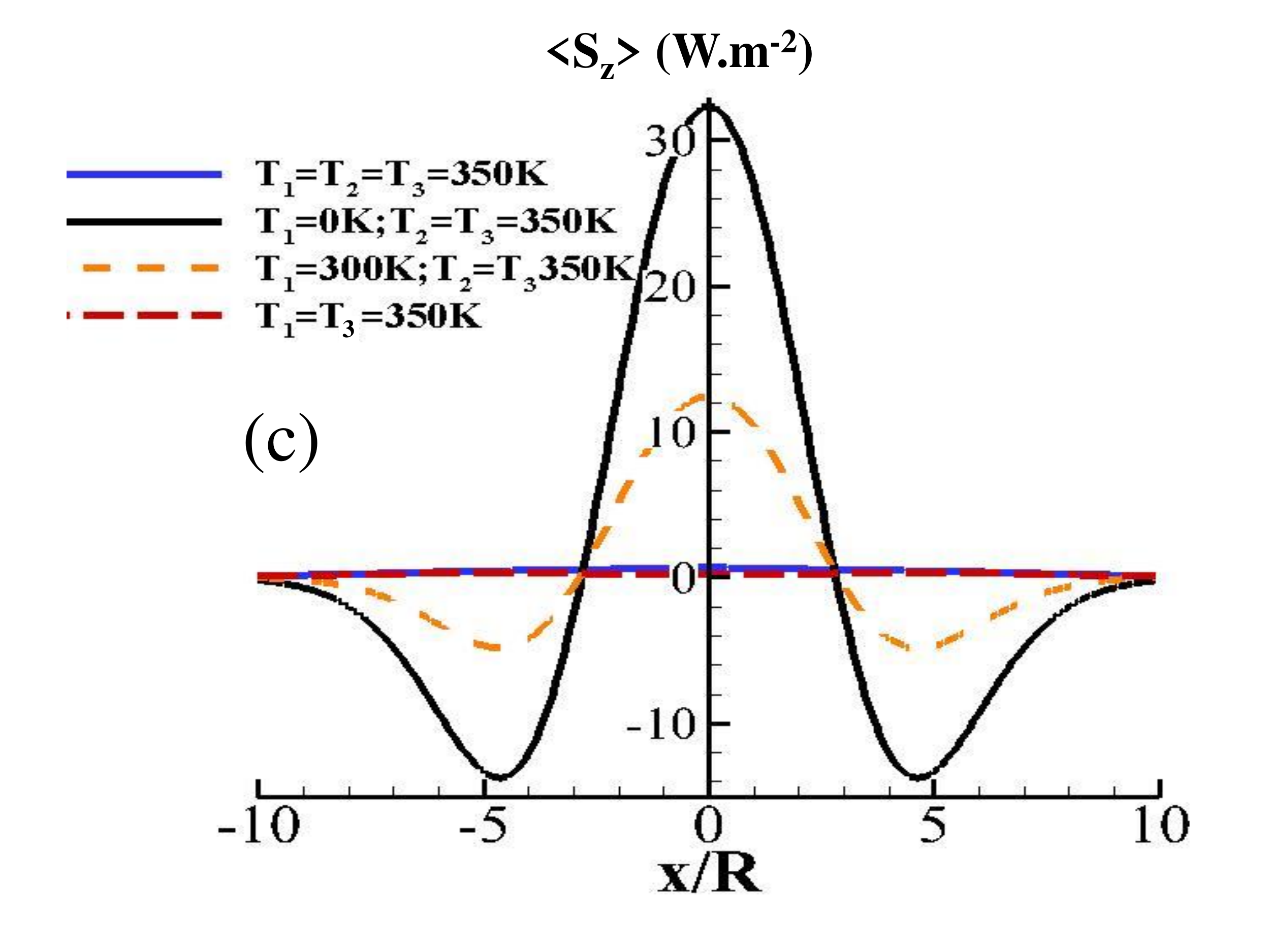}
\caption{(a) Normal component $<S_z>$ of Poynting vector radiated through the surface $z=0$ by a three-tips SThM setup made with three glass emitters held at $T_2=T_3=350K$ (red) and $T_1=300K$ (blue) for different angular opening ($R=20 nm$, $h=4R$ and $l=5R$). The inset shows the flux at $z=0$ for a single particle at $T=350K$ (red) and $T=300K$ (blue) at $z=4R$ (solid) and $z=4R+l cos(80°)$ (dashed) (b) Magnitude of Poynting vector and vector field in the $(x,z)$ plane radiated by a multi-tip setup of angular opening  $\theta=80^{\circ}$. (c) Normal component $<S_z>$ of Poynting vector at $z=0$ with respect to the temperature differences. The long dashed red line corresponds to the flux emitted by two particles (2 and 3) at the same temperature.
\label{focus_diff_temp}}
\end{figure}

\begin{equation}
<\widetilde{S}_{\zeta}(\bold{r},\omega)>=2\frac{\omega^2}{c^2} \underset{i=1}{\overset{N}{\sum}}a_{\zeta}(\bold{r},\bold{r_i},\omega)\Theta(T(\bold{r_i}),\omega),
\label{Poynting_bis}
\end{equation}
where $\Theta(T,\omega)={\hbar\omega}/[{e^{\frac{\hbar\omega}{k_B T}}-1}]$ is the mean energy of a harmonic oscillator at temperature $T$ 
and
\begin{equation}
a_{\zeta}(\bold{r},\bold{r_i},\omega)\equiv\epsilon_{\zeta\gamma\beta}Im\{\mathds{G}^{EE}_{\beta\eta}(\bold{r},\bold{r_i})\mathds{G}^{HE*}_{\gamma\eta}(\bold{r},\bold{r_i})\}Im(\alpha_i)
\label{kernel}
\end{equation}
with summation over the repeated index $\gamma$, $\beta$ and $\eta$.

To discuss the specificities of the multi-tip near-field thermal microscopy let  us consider a triangular three-tip configuration as depicted in the inset of Fig.2-a where all  emitters (nanospheres glass) are placed at the apex of an isosceles triangle and are thermalized at the same temperature. We also neglect  the far-field heat flux exchanged with the external bath in front of near-field heat flux radiated by the multi-tip setup  in its close neighborhood~\cite{Yannopapas}. This is equivalent to assume an environnement at $0\:K$. The normal component $<S_z>=\int\frac{d\omega}{2\pi}<\widetilde{S}_z(\omega)>$ of Poynting vector radiated in near-field by this system across the plan $z=0$ at a distance $h=4R$ from the lowest apex is shown in Fg.2-a for different angular opening $\theta$ and compared with the flux radiated by a single emitter at the same distance and same temperature. We consider the system in steady state regime that is after a thermalization time which is typically, for many-body systems, of the order of the millisecond~\cite{Riccardo}. With this thermalized system the FWHM of heat flux trivially increases with $\theta$ and it is always greater than with a single emitter. This behavior is simply due to the fact that the emission pattern of the whole structure corresponds to the superposition of flux radiated  by three fluctuating dipoles with the same distribution function and with a polarizability equals to its dressed polarizability to take into account the cooperative effects. In near-field regime we see  (Fig.2(b)) that the flux lines diverge from the source in a similar way as around a single object.
In far-field regime (i.e. $\underset{i}{Min}\mid\mathbf{r}-\mathbf{r_{i}}\mid\gg \lambda_{th}$ ) the thermal emission can  be calculated using the Kirchoff's law from the absorption cross-section of the system which reduces after a straightforward calculation  using the Landauer formalism~\cite{pbaPRB2016} to
\begin{equation}
\sigma_a(\omega)=4\pi A(\bold{u})\mid\underset{i=1}{\overset{N}{\sum}}\bold{a}(\bold{r},\bold{r_i},\omega).\bold{u}\mid
\label{absorption}
\end{equation}
$A(\bold{u})$ being the apparent surface of system in the direction $\bold{u}$ along which it is shined.

On the contrary, for a system out of thermal equilibrium, the situation radically changes. In particular the Kirchoff's law fails to describe its thermal behavior in far-field regime since according to expression~(\ref{Poynting_bis}) light emission and therefore light absorption is not anymore an intrinsic property of the system but it closely depends on its local thermal state, that is on its temperature.  Hence we see in  Figs. 3 by heating up the base of triangular system  that the heat flux can be focused  on a much smaller spot than with a fully thermalized system (Fig.2). In this case the FWHM of heat flux profile at $z=0$ can even be between $1.5$ to $2$ time smaller than with a single tip depending on the angular opening of multi-tip setup. It is northwhile to note that  the heat flux is enhanced by approximately two orders of magnitude compared to a  thermalized  system of 3-dipoles or a single  emitter when the triangle base is heated up by only $50K$. 
\begin{figure}%[h!]
\includegraphics[scale=0.3]{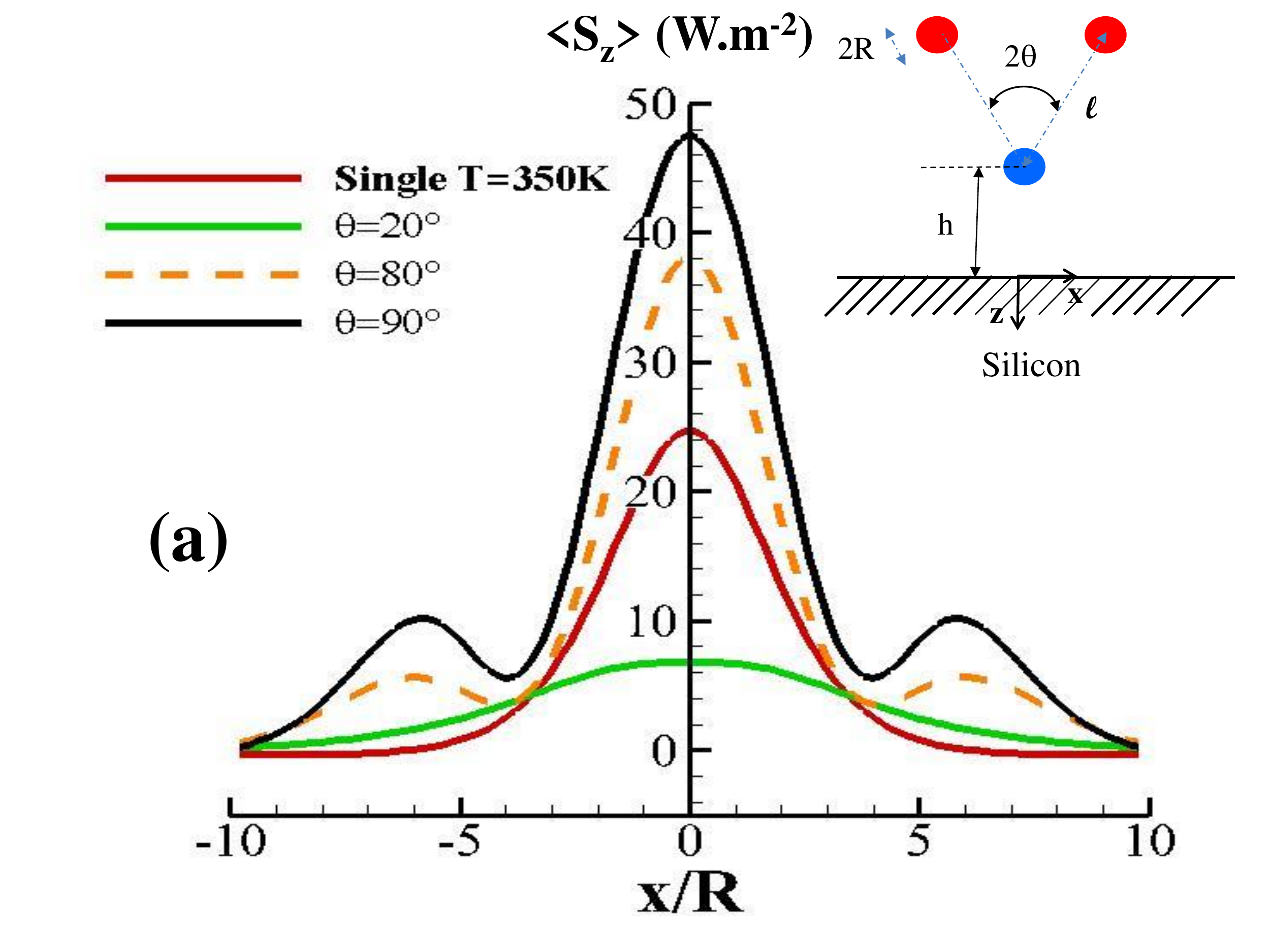}
\includegraphics[scale=0.3]{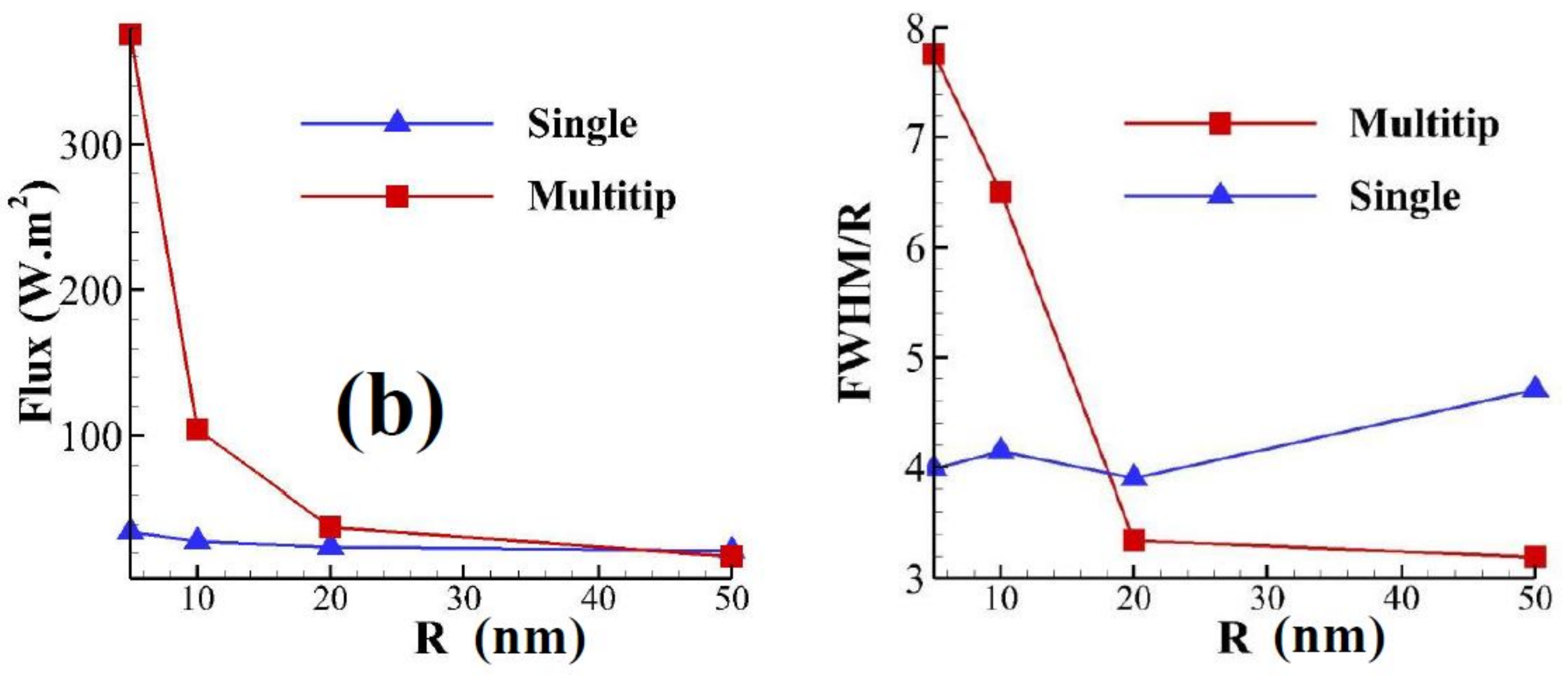}
\includegraphics[scale=0.3]{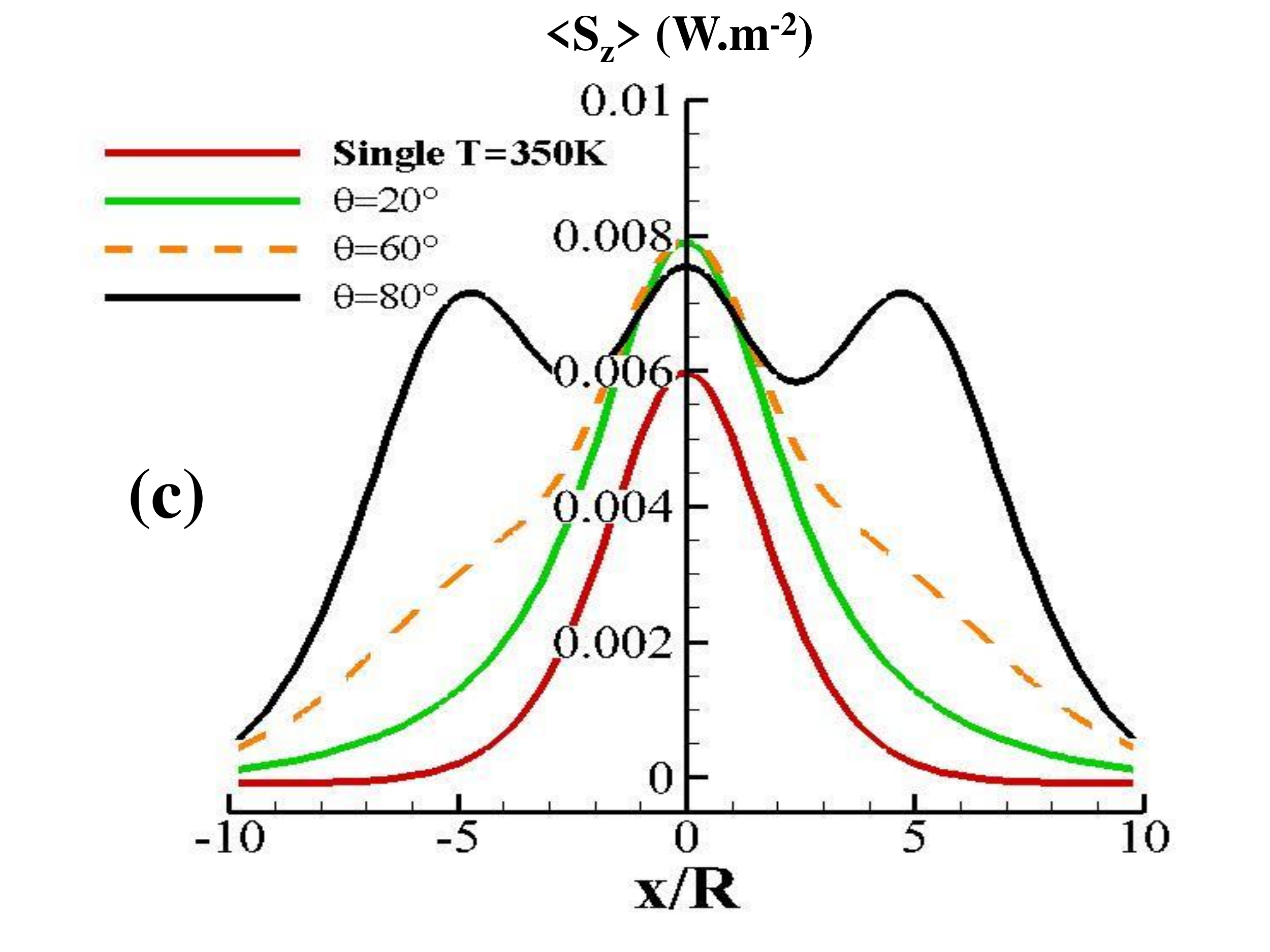}
\caption{ Normal component $<S_z>$ of Poynting vector radiated through the surface $z=0$ by a three (a) glass nanoparticles above a silicon sample  (same parameters as in Fig. 3). The red curve show the flux at $z=0$ for a single particle at $T=350K$ located at $z=4R$ above the silicon sample. (b) $<S_z>$ at $z=0$ and $x=0$ and FWHM for different radius with an angular opening of $\theta=80^{\circ}$. (c)  Same as (a) with gold nanoparticles.
\label{focus_diff_temp_interface}}
\end{figure}

We also observe  in Fig.3(a) that for systems with a small angular opening  the heat flux can back propagates toward the emitting system itself just below the cold tip while it flows toward the surrounding environment  around this region, showing so that a multi-tip setup can be used as a local heat pump. On the other hand, for large angular openings this pumping effect is off-center. The mapping of Poynting vector field (Fig. 3(b) ) in the $(x,z)$ plane shows that this effect results from the local bending of flux lines. The mechanism responsible for the redirection of energy flow toward the system is directly related, according to expression~(\ref{Poynting_bis}), to the presence of a local temperature difference inside the system. The consequences of this gradient on the heat pumping and focusing effects  are highlighted in Fig.3(c) in the case of a triangular system subject to  two temperature differences between its apex and its base. We see that the magnitude of those effects depends directly on this temperature discrepancy. In particular we observe that while this  magnitude  increases when the thermal emission of the apex particle is reduced the presence of this  particle is nevertheless fundamental for these effects to exist. Indeed we see in Fig. 3(c) that without this particle the magnitude of heat flux falls down by approximateley two orders of magnitude. Hence, even if the apex particle is non-emitting ($T=0\:K$) it participates both to the local heat funneling (heat pumping and heat focusing) by distording the flux lines and to the exaltation by dressing effect of the thermal emission in near-field regime of two others particles.

The above analysis holds when the overall  system is in vacuum. In the following we discuss how the highlighted effects are impacted by the presence of an interface. To do so we consider similar dipolar systems held close to the ambiant temperature which are brought, at subwavelength distances, from a simple silicon (Si) sample. In the spectral range where most of heat exchanges take place the dielectric permittivity of Si is close to $\epsilon_{Si}=11.5$ ~\cite{Palik98}. In Fig.4 we show the heat flux radiated by a triangular system made of  glass nanoparticles (Fig. 4(a)) and of gold nanoparticles (Fig. 4(c)), respectively. The results show that, for polar particles, the heat flux is amplified by more than a factor when the particles are parallel to the surface compared to the same configuration in vacuum. The flux is even amplified by almost a factor 4  for an angular opening of $\theta=80^{\circ}$. On the other hand for metallic particles (Fig.4(c)) the amplification factor remains smaller than two highlighting so the important role play by the resonant modes (surface phonon-polaritons) supported by the glass particles.  As  the focusing effect is concerned it is slightly  improved by the presence of the interface when $\theta=80^{\circ}$. Indeed, in this case the $FWHM=3.34R$ while for a single particle $FWHM_{single}=3.9R$. In Fig. 4(b) we see that this focusing evolves non monotically with the nanoparticles radius. It is worthwhile to note that the FWHM for a multitip setup is always smaller than with a single tip when $R>20 nm$ (for large particles the contribution of magnetic moment should also be taken into account and would required further investigations) . For small particles the FWHM  can be comparable to the superparamagnetic limit of usual recording materials~\cite{super1,super2}. In this case the focusing is not necessary better than with a single particle but the heat flux can be ten times more important than with a single emitter.  However, the optimization of  focusing effect with respect to the geometrical configuration, the materials properties and the temperature distribution remains today an open problem which goes far beyond the scope of the present Letter.

In summary, we have introduced the concept of multi-tip near-field scanning thermal microscopy and demonstrated that it  can be used to tailor heat flux at nanoscale. We have shown that this flux can be focused and even amplifield  in tiny regions which are much smaller than the diffraction limit and even smaller than with a single tip near-field SThM. The relatively simple model used to describe the tips does not allow to setups with tips close to the contact. For such separation distances the dipolar approximation fails du describe properly the thermal behaviour of emitters and the high orders modes have to be taken into account. These modes give a supplementary degree of freedom to sculpt the field radiated by the sources and we can expect even better performances with such configurations.
The significance of multi-itp  scanning thermal microscopy to analyze at nanoscale the thermal state of solid surfaces is obvious. It could also find broad applications  in many other fields to measure for instance thermal properties such as  heat capacity and the infrared spectra of nano-objects.

\begin{acknowledgments}
The author acknowledges discussions with Riccardo Messina and thanks  Mondher Besbes for his help on graphical representations.
\end{acknowledgments}

\end{document}